\documentclass{emulateapj}

\newcommand{\Kepler}{{\it Kepler}}
\newcommand{\Spitzer}{{\it Spitzer}}
\newcommand{\hipparcos}{{\it Hippacos}}
\newcommand{\vespa}{\texttt{vespa}}

\newcommand{\be}{\begin{equation}}
\newcommand{\ee}{\end{equation}}

\newcommand{\metallicity}{[M/H]}
\newcommand{\msun}{M$_\odot$}
\newcommand{\rsun}{R$_\odot$}

\newcommand{\diststar}{116 $\pm$ 18 pc}

\newcommand{\kms}{\ensuremath{\rm km\,s^{-1}}}
\newcommand{\ms}{\ensuremath{\rm m\,s^{-1}}}

\newcommand{\teff}{6199}
\newcommand{\teffe}{50}

\newcommand{\loggspc}{4.13}
\newcommand{\loggespc}{0.1}

\newcommand{\logg}{4.18}
\newcommand{\logge}{0.1}

\newcommand{\mh}{-0.11}
\newcommand{\mhe}{0.08}

\newcommand{\vsini}{7.13}
\newcommand{\vsinie}{0.5}

\newcommand{\mstar}{1.15}
\newcommand{\mstare}{0.064}

\newcommand{\rstar}{1.4}
\newcommand{\rstare}{0.19}

\newcommand{\rearth}{R$_\oplus$}

\newcommand{\perpld}{$156\substack{+163 \\ -78}$}%
\newcommand{\perple}{$131\substack{+61 \\ -36}$}%
\newcommand{\perplf}{$324\substack{+121 \\ -126}$}%

\newcommand{\ldone}{0.311}
\newcommand{\uldone}{0.048}
\newcommand{\ldtwo}{0.31}
\newcommand{\uldtwo}{0.13}
\newcommand{\rprstb}{0.0188}
\newcommand{\urprstb}{0.0011}
\newcommand{\arstb}{19.5}
\newcommand{\uarstb}{4.5}
\newcommand{\inclb}{88.4}
\newcommand{\uinclb}{1.6}
\newcommand{\impb}{0.55}
\newcommand{\uimpb}{0.28}
\newcommand{\rplb}{2.90}
\newcommand{\urplb}{0.44}
\newcommand{\perplb}{15.5712}
\newcommand{\uperplb}{0.0012}
\newcommand{\ttransitb}{2457152.2844}
\newcommand{\uttransitb}{0.0021}
\newcommand{\rprstc}{0.0166}
\newcommand{\urprstc}{0.0012}
\newcommand{\arstc}{73}
\newcommand{\uarstc}{18}
\newcommand{\inclc}{89.58}
\newcommand{\uinclc}{0.52}
\newcommand{\impc}{0.53}
\newcommand{\uimpc}{0.29}
\newcommand{\rplc}{2.56}
\newcommand{\urplc}{0.40}
\newcommand{\perplc}{31.6978}
\newcommand{\uperplc}{0.0040}
\newcommand{\ttransitc}{2457163.1659}
\newcommand{\uttransitc}{0.0027}
\newcommand{\rprstd}{0.0259}
\newcommand{\urprstd}{0.0015}
\newcommand{\tdurd}{12.71}
\newcommand{\utdurd}{0.26}
\newcommand{\impd}{0.50}
\newcommand{\uimpd}{0.27}
\newcommand{\rpld}{3.96}
\newcommand{\urpld}{0.59}
\newcommand{\ttransitd}{2457166.2629}
\newcommand{\uttransitd}{0.0016}
\newcommand{\rprste}{0.03613}
\newcommand{\urprste}{0.00096}
\newcommand{\tdure}{13.007}
\newcommand{\utdure}{0.088}
\newcommand{\impe}{0.31}
\newcommand{\uimpe}{0.17}
\newcommand{\rple}{5.51}
\newcommand{\urple}{0.77}
\newcommand{\ttransite}{2457142.01656}
\newcommand{\uttransite}{0.00076}
\newcommand{\rprstf}{0.0672}
\newcommand{\urprstf}{0.0013}
\newcommand{\tdurf}{18.998}
\newcommand{\utdurf}{0.051}
\newcommand{\impf}{0.227}
\newcommand{\uimpf}{0.089}
\newcommand{\rplf}{10.2}
\newcommand{\urplf}{1.4}
\newcommand{\ttransitf}{2457186.91451}
\newcommand{\uttransitf}{0.00032}

\newcommand{\thisstar}{HIP~41378}
\newcommand{\thisfirstplanet}{HIP~41378~b}
\newcommand{\thissecondplanet}{HIP~41378~c}
\newcommand{\thisthirdplanet}{HIP~41378~d}
\newcommand{\thisfourthplanet}{HIP~41378~e}
\newcommand{\thisfifthplanet}{HIP~41378~f}

\usepackage{xcolor}
\usepackage{hyperref}
\usepackage{breakurl}
\usepackage{amsmath}
\usepackage{graphicx}
\usepackage{verbatim}
\usepackage{booktabs}

\slugcomment{}

\shorttitle{Five Planets Transiting a Bright Star}
\shortauthors{Vanderburg et al.}

\begin{document}

\title{Five Planets Transiting a Ninth Magnitude Star}
\author{Andrew Vanderburg\altaffilmark{1,$\diamondsuit$,$\spadesuit$}, Juliette C. Becker\altaffilmark{2,$\diamondsuit$}, Martti H. Kristiansen\altaffilmark{3,4}, Allyson Bieryla\altaffilmark{1}, Dmitry A. Duev\altaffilmark{5}, Rebecca Jensen-Clem\altaffilmark{5}, Timothy D. Morton\altaffilmark{6}, David W. Latham\altaffilmark{1}, Fred C. Adams\altaffilmark{2,7}, Christoph Baranec\altaffilmark{8}, Perry Berlind\altaffilmark{1}, Michael L. Calkins\altaffilmark{1}, Gilbert A. Esquerdo\altaffilmark{1}, Shrinivas Kulkarni\altaffilmark{5},  Nicholas M. Law\altaffilmark{9}, Reed Riddle\altaffilmark{5}, Ma\"{i}ssa Salama\altaffilmark{10}, \& Allan R. Schmitt\altaffilmark{11} }
 
\altaffiltext{1}{Harvard--Smithsonian Center for Astrophysics, 60 Garden St., Cambridge, MA 02138}
\altaffiltext{2}{Astronomy Department, University of Michigan, Ann Arbor, MI 48109, USA}
\altaffiltext{3}{DTU Space, National Space Institute, Technical University of Denmark, Elektrovej 327, DK-2800 Lyngby, Denmark}
\altaffiltext{4}{Brorfelde Observatory, Observator Gyldenkernes Vej 7, DK-4340 T\o{}ll\o{}se, Denmark}
\altaffiltext{5}{California Institute of Technology, Pasadena, CA, 91125, USA}

\altaffiltext{6}{Department of Astrophysical Sciences, 4 Ivy Lane, Peyton Hall, Princeton University, Princeton, NJ 08544, USA}
\altaffiltext{7}{Physics Department, University of Michigan, Ann Arbor, MI 48109, USA}
\altaffiltext{8}{University of Hawai`i at M\={a}noa, Hilo, HI 96720, USA}
\altaffiltext{9}{University of North Carolina at Chapel Hill, Chapel Hill, NC, 27599, USA}

\altaffiltext{10}{University of Hawai`i at M\={a}noa, Honolulu, HI 96822, USA}
\altaffiltext{11}{Citizen Scientist}
\altaffiltext{$\diamondsuit$}{NSF Graduate Research Fellow}
\altaffiltext{$\spadesuit$}{\url{avanderburg@cfa.harvard.edu}}

\begin{abstract}
The \Kepler\ mission has revealed a great diversity of planetary systems and architectures, but most of the planets discovered by \Kepler\ orbit faint stars. Using new data from the K2 mission, we present the discovery of a five planet system transiting a bright (V = 8.9, K = 7.7) star called \thisstar. \thisstar\ is a slightly metal-poor late F-type star with moderate rotation ($v\sin i \simeq$ 7 \kms) and lies at a distance of \diststar\ from Earth. We find that \thisstar\ hosts two sub-Neptune sized planets orbiting 3.5\% outside a 2:1 period commensurability in 15.6 and 31.7 day orbits. In addition, we detect three planets which each transit once during the 75 days spanned by K2 observations. One planet is Neptune sized in a likely $\sim$ 160 day orbit, one is sub-Saturn sized likely in a $\sim$ 130 day orbit, and one is a Jupiter sized planet in a likely $\sim$ 1 year orbit. We show that these estimates for the orbital periods can be made more precise by taking into account dynamical stability considerations. We also calculate the distribution of stellar reflex velocities expected for this system, and show that it provides a good target for future radial velocity observations. If a precise orbital period can be determined for the outer Jovian planet through future observations, it will be an excellent candidate for follow-up transit observations to study its atmosphere and measure its oblateness. 

\end{abstract}


\section{Introduction}

The \Kepler\ spacecraft (launched in 2009) has been a tremendously successful planet hunter \citep{borucki, koi2, koch}. Over the course of its four year original mission, \Kepler\ discovered thousands of planetary candidates around distant stars \citep{coughlin}, demonstrating the diversity and prevalence of planetary systems \citep[e.g.][]{muirhead, fabrycky, kepler47, mortonswift}. \Kepler's contributions include measuring the size distribution of exoplanets \citep{howard, fressin, petigura1}, understanding the composition of planets intermediate in size between the Earth and Neptune \citep{weissmarcy, angie}, measuring the prevalence of rocky planets in their host star's habitable zones \citep{dc13, dc15, petigura2, foreman-mackey, burke}, and uncovering the wide range of orbital architectures like tightly packed planetary systems \citep{campante}, and planets in and (more commonly) near low order mean motion resonances \citep{carter, steffen}. 

While \Kepler's discoveries have been illuminating, many questions about these systems still remain, and are difficult to answer because \Kepler\ planet host stars are typically faint. \Kepler\ observed a single 110 square degree field for four years, and that deep survey strategy produced many candidates too faint for intensive follow-up observations. Because of this limitation, our understanding of the properties of \Kepler\ planets is incomplete. For instance, radial velocity (RV) follow-up of the brightest planet candidates on short period orbits have found that planets smaller than about 1.5 earth radii are typically rocky \citep{rogers, dressing}, but masses measured from the inversion of transit-timing variations (TTVs) show a population of low density planets orbiting farther from their host stars than most transiting planets with RV measurements \citep{steffenpopulation}. There are few planets with masses measured from TTVs transiting stars bright enough for RV followup, and in the few cases where both types of measurements are possible, there is not always perfect agreement \citep[e.g. KOI 94:][]{koi94weiss, koi94masuda}. \Kepler\ has also discovered giant planets transiting stars on long period orbits \citep{kippingsnowline, kippingjupiteranalog} whose atmospheres could be studied via transmission spectroscopy \citep[][]{dalba}, but these planets orbit stars fainter than most stars hosting planets with well characterized atmospheres \citep[e.g.][]{deming, kreidberg}. 

The end of the original Kepler mission in 2013 due to a mechanical failure has led to new opportunities for the \Kepler\ spacecraft to discover planets orbiting brighter stars than before. In its new K2 extended mission \citep{howell}, Kepler observes many fields along the ecliptic plane for up to 80 days. Over the course of the K2 mission, \Kepler\ could survey up to 20 times the sky area as it did in its original mission, greatly increasing the number of bright stars and other rare objects observed. The K2 mission has already yielded transiting planets and candidates around (for example) stars as bright as 8th magnitude \citep{v15}, nearby M-dwarfs \citep{ crossfield,petigura, hirano,schlieder}, and stars in nearby open clusters \citep{mann}.

\begin{figure*}[t!] 
   \centering
   \includegraphics[width=6.5in]{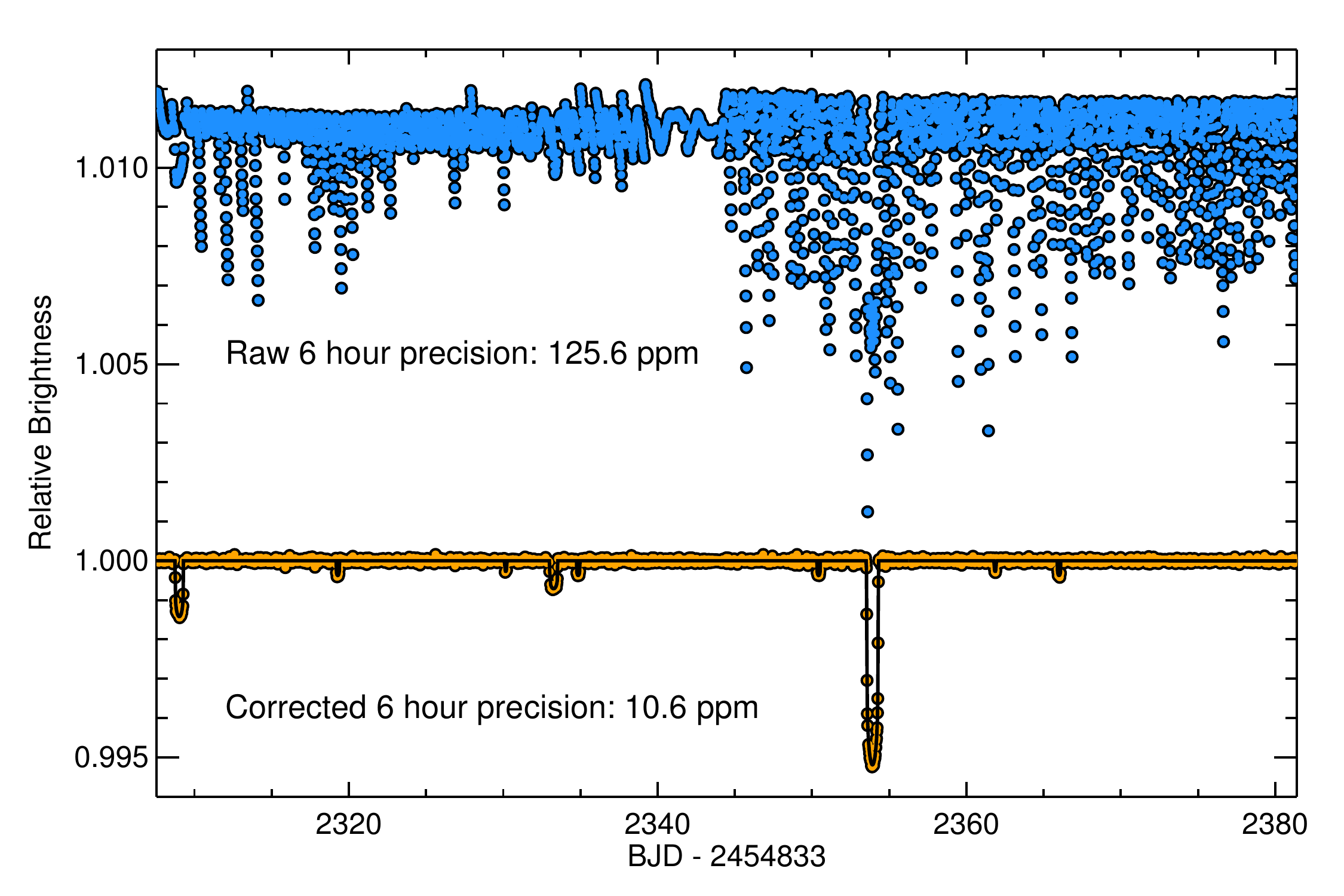} 
   \caption{Raw K2 light curve (blue, top), and systematic corrected light curve (orange, bottom). The best-fit transit model is shown as a black line over the orange systematics-corrected light curve. We have flattened the light curve by removing our best-fit long term trend from the simultaneous transit/systematic fit. The three deepest transits are single-events, and are highly inconsistent with each other in terms of depth and duration. The systematics correction improves the photometric precision to a level of 10 ppm per six hours -- comparable to the best light curves from the original \Kepler\ mission.}
   \label{lc}
\end{figure*}

Because it only observes stars for about 80 days before moving onto new fields, K2 is not as sensitive to planetary systems with complex architectures as the original 
\Kepler\ mission. While \Kepler\ detected systems with up to seven transiting planets \citep{schmitt, cabrera}, K2 has not yet discovered any systems with more than three transiting planet candidates\footnote{The WASP-47 system hosts four planets, only three of which are known to transit \citep{wasp47, becker, vanmalle}.} \citep{sinukoff, v15}. 

In this paper, we report the discovery of a system of five transiting planets using K2 data. The host star is one of the brightest planet host stars from either \Kepler\ or K2 with a V magnitude of 8.9 and a K magnitude of 7.7, and has a trigonometric parallax based distance of \diststar. The planetary system displays a rich architecture, with two sub-Neptunes slightly outside of mean motion resonance and three larger planets in longer period orbits. The outer planet is a gas giant on an approximately 325 day orbit (detected by a single transit in the 80 days of K2 data), and if a precise orbital period can be recovered, will be a favorable target for follow-up observations. In Section \ref{observations} we describe both the K2 observations and our follow-up observations taken to characterize the system and rule out false-positive scenarios. Section \ref{analysis} presents our analysis and determination of planet parameters, our statistical validation of the transit signals as genuine planets, and includes dynamical constraints requiring system stability. A discussion of our results is presented in Section \ref{discussion}, followed by a summary in Section \ref{summary}.

\section{Observations}\label{observations}

\subsection{K2 Light Curve}\label{lightcurve}

\thisstar, also known as EPIC 211311380, was observed by the \Kepler\ space telescope during Campaign 5 of its extended K2 mission for a period of about 75 days between 27 April 2015 and 10 July 2015.  We downloaded the calibrated target pixel files for \thisstar\ from the Mikulski Archive for Space Telescopes, and processed the light curve following \citet{vj14} and \citet{v15} to produce a photometric light curve and remove systematic effects from the light curve caused by the K2 mission's unstable pointing. Visual inspection of the K2 light curve by one of us (M.H.K.) revealed the presence of nine individual transit events at high confidence. A Box Least Squares periodogram search \citep{kovacs} of the light curve revealed that four of the nine individual transits occur periodically every 15.57 days. Two other transits are also consistent in shape, duration, and depth and occur approximately 31.7 days apart, suggesting two planets near the 2:1 mean motion resonance (MMR). The remaining three transit events seen in the K2 light curve are not consistent with one another and each have longer durations than the repeating transits, suggesting orbital periods longer than the 75 days of K2 observations. We designate five planet candidates around \thisstar; we refer to the inner two candidates as \thisfirstplanet\ and \thissecondplanet, in order of increasing orbital period, and we refer to the outer three planet candidates as \thisthirdplanet, \thisfourthplanet, and \thisfifthplanet, in order of increasing transit duration. 

After identifying transits in the light curve, we reprocessed the K2 data by simultaneously fitting the transits with the K2 roll systematics and the long-term variability in the star's light curve, as described in \citet{v15}. The resulting light curve has a precision of 10.6 parts per million (ppm) per six hours or 38 ppm per 30 minute long-cadence exposure and is shown in Figure \ref{lc} along with the raw uncorrected light curve. 

The photometric aperture we used to extract the K2 light curve is large due to \thisstar's brightness. We inspected archival imaging from the Palomar Observatory Sky Survey and found that another nearby star about five magnitudes fainter than \thisstar\ falls inside our photometric aperture. We checked that the transit signals are in fact centered on \thisstar\ by extracting a light curve from smaller photometric apertures which exclude the nearby star. Although the photometric precision of the light curves from smaller apertures are significantly worse than the photometric precision of our original large aperture, we detect all nine transits at the same depths as the original light curve. We therefore conclude the transits are not centered on the fainter star.

\subsection{High Resolution Spectroscopy}

We observed \thisstar\ with the Tillinghast Reflector Echelle Spectrograph (TRES) on the 1.5 meter telescope at Fred L. Whipple Observatory on Mt. Hopkins, Arizona. We obtained spectra on four different nights in January and February 2016. The spectra were obtained at a spectral resolving power of $\lambda/\Delta\lambda$ = 44,000, and exposures of 360 - 450 seconds yielded spectra with signal-to-noise ratios of 90 to 110 per resolution element. We see no evidence for chromospheric calcium II emission from the H-line at 396.85 nm. We cross correlated the four spectra with a model spectrum and inspected the resulting cross correlation functions (CCFs). There is no evidence in the CCFs for additional second sets of stellar lines. We measure an absolute radial velocity for \thisstar\ of 50.7 \kms, and the four individual spectra show no evidence for high-amplitude radial velocity variations. We measured relative radial velocities by cross correlating each observation with the strongest observation and found no evidence for RV variations greater than TRES's intrinsic RV precision of 15 \ms.  

\subsection{Adaptive Optics Imaging}

We observed \thisstar\ with the Robo-AO adaptive optics (AO) system on the 2.1 meter telescope at the Kitt Peak National Observatory \citep{baranec, lawroboao,roboaokp}. Robo-AO is a robotic laser guide star adaptive optics system, which has recently moved to the 2.1 meter telescope at Kitt Peak from the 1.5 m telescope at Palomar Observatory. We obtained an image on 2 April 2016 with an $i'$-band filter. The observation consisted of a series of exposures taken at a frequency of 8.6 Hz, which were then shifted and added using \thisstar\ as the tip--tilt guide star. The total integration time was 120 seconds. 

The resulting image showed no evidence for any companions to \thisstar\ within the 36$\arcsec\times$36$\arcsec$ Robo-AO field of view; the nearby star discussed Section \ref{lightcurve} falls outside the field of view. The AO observations allow us to exclude the presence of companion stars two magnitudes fainter than \thisstar\ at a distance of 0$\farcs$25, and stars four magnitudes fainter at a distance of 0$\farcs$7 with 5-$\sigma$ confidence. 

\section{Analysis}\label{analysis}

\subsection{Spectroscopic and Stellar Properties}
\label{sec:stellar_prop}
We measured spectroscopic properties from each of the four TRES observations using the Stellar Parameter Classification \citep[SPC, ][]{buchhave, buchhave14} method. SPC cross-correlates observed spectra with a suite of synthetic spectra based on \citet{kurucz} atmosphere models, and interpolates the CCFs to determine the stellar effective temperature, metallicity, surface gravity, and projected rotational velocity. The four exposures yield consistent spectroscopic parameters, which are summarized in Table \ref{bigtable}.  \thisstar\ appears to be a slightly evolved late F-type star with a temperature of \teff\ $\pm$ \teffe\ K, a surface gravity of $\log{g_{cgs,\rm{~SPC}}}$ = \loggspc\  $\pm$ \loggespc, a metallicity [M/H] of \mh\ $\pm$ \mhe, and a projected rotation velocity of \vsini\  $\pm$ \vsinie\ \kms. 

We determined the mass and radius of \thisstar\ using an online interface\footnote{\url{http://stev.oapd.inaf.it/cgi-bin/param}} to interpolate the star's temperature, metallicity, V-band magnitude, and \hipparcos\ parallax onto Padova stellar evolution tracks, as described by \citet{dasilva}. We find that \thisstar\ has a mass of \mstar\ $\pm$ \mstare\ \msun\ and a radius of \rstar\ $\pm$ \rstare\ \rsun. The models predict a slightly stronger surface gravity of $\log{g_{cgs}}$ = \logg\ $\pm$ \logge\ than the spectroscopic measurement of \loggspc\  $\pm$ \loggespc, but the surface gravities are consistent at the 1--$\sigma$ level. The consistency between the spectroscopic and model surface gravities provide an independent check that our stellar parameters are reasonable. 

\subsection{Transit Analysis}
\label{sec:lc_analysis}
We analyzed the K2 light curve by simultaneously fitting the five transiting planet candidates and a model for low frequency variability using a Markov Chain Monte Carlo algorithm with an affine invariant ensemble sampler \citep{goodman}. We fit the five transiting planet candidates with \citet{mandelagol} transit models, and we modeled the low frequency variations with a basis spline. For the two inner candidates, we fit for the orbital period, time of transit, scaled semi-major axis ($a/R_{\star}$), orbital inclination, and planet to star radius ratio ($R_{P}/R_{\star}$). For the three outer candidates with only one transit, we fit for the transit time, duration (from the first to fourth contact), transit impact parameter, and planet to star radius ratio. We fit for quadratic limb darkening coefficients for all five transits simultaneously, using the the $q_1$ and $q_2$ parametrization from \citet{kippingld}. We imposed no priors on these parameters other than requiring impact parameters be positive. We accounted for the effects of the \Kepler\ long cadence exposure time by oversampling each model data point by a factor of 30 and performing a trapezoidal numerical integration. We did not account for any asymmetry in the transit light curve due to eccentricity -- this effect scales with $(a/R_{\star})^{-3}$ and is too small to detect for long period planets like these \citep{winn}. We note that our choice to parameterize the orbits by their inclinations is an approximation -- although orbits are uniformly distributed in $\cos{i}$, not $i$, the difference is negligible for nearly edge on orbits like those of the planets transiting \thisstar. We performed a Monte Carlo calculation and found that the different parameterizations only change our final measured inclinations by roughly $10^{-4}$ degrees, much less than our measured uncertainties in inclination.

We sampled the parameter space using 150 walkers evolved through 40,000 links, and removed the first 20,000 links during which time the chains were ``burning-in'' to a converged state. This yielded a total of 3,000,000 individual samples. We tested the convergence of the MCMC chains by calculating the Gelman-Rubin statistic \citep{gelmanrubin}. For each parameter, the Gelman-Rubin statistic was below 1.04, indicating our MCMC fits were well converged. 

\begin{figure*}[t!] 
   \centering
   \includegraphics[width=6.5in]{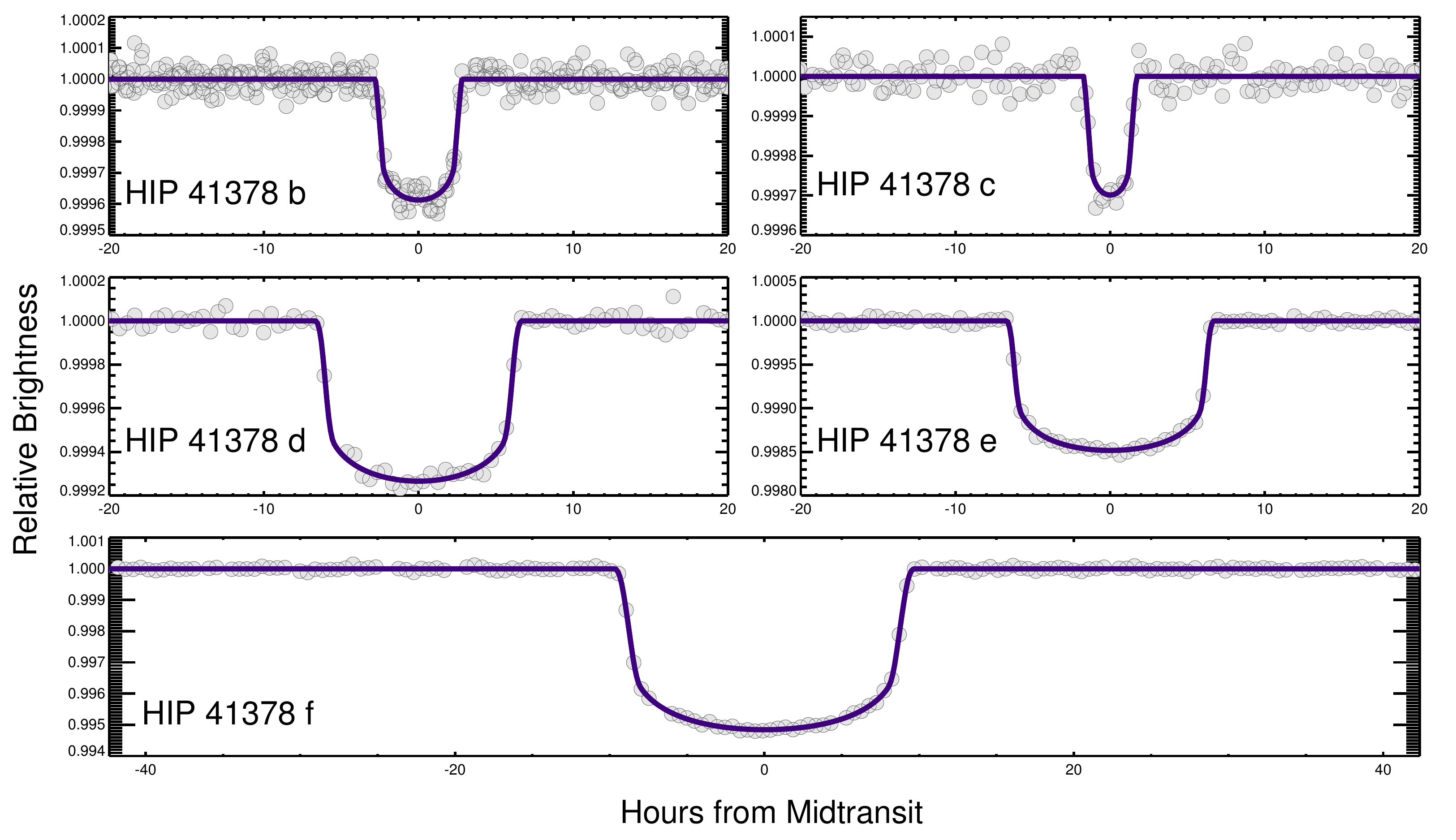} 
   \caption{Phase-folded light curve for each of the five transiting planets in the \thisstar\ system. The individual K2 long cadence data points are shown as grey circles, and the best-fit transit model is shown as a thick purple line. The scaling on the x-axis is the same for each sub-panel. In each panel, we have subtracted the best-fit transit model for the other four planets for clarity. }
   \label{transitfit}
\end{figure*}

We plot the transit light curves for each planet and the best-fitting transit model in Figure \ref{transitfit}. 

\subsection{Statistical Validation}\label{validation}

The transit signals in the K2 light curve of \thisstar\ that we attribute to transiting planets could in principle have a non-planetary astrophysical origin. In this subsection, we argue that astrophysical false positive scenarios are unlikely in the case of the \thisstar\ system, and a planetary interpretation of the transit signals is well justified. 

We began by calculating the false positive probability (FPP) of the inner two planet candidates, which both have precisely measured orbital periods from multiple transits in the K2 light curve, using the \vespa\ software package \citep{morton2012, morton2015}. \texttt{Vespa} takes information about the transit shape, orbital period, host star parameters, location in the sky, and observational constraints and calculates the likelihood that a given transit signal has an astrophysical origin other than a transiting planet.  We used \vespa\ to calculate the FPP of \thisfirstplanet\ and \thissecondplanet\ given constraints on the depth of any secondary eclipse from the K2 light curve and limits on any nearby companion stars in the K2 aperture from Robo-AO. We also used the fact that our radial velocity measurements of \thisstar\ show no variations greater than about 15 \ms\ to exclude all foreground eclipsing binary false positive scenarios. Given these constraints, we calculate that the FPP for \thisfirstplanet\ is very small, of order $2\times 10^{-6}$, and the FPP of \thissecondplanet\ is $3\times10^{-3}$, somewhat larger but still quite low. These FPPs do not take into account the fact that we detect five different candidate transit signals towards \thisstar\ and that the vast majority of \Kepler\ multi-transiting candidate systems are real planetary systems \citep{latham, lissauer}. \citet{lissauer} estimate that being in a system of three or more candidates increases the likelihood of a given transit signal being real by a factor of $\sim$ 50-100. Taking this multiplicity argument into account, the FPP for \thisfirstplanet\ decreases to roughly $10^{-7}$ and the FPP for \thissecondplanet\ decreases to roughly $10^{-4}$. We therefore consider \thisfirstplanet\ and \thissecondplanet\ to be validated as genuine planets.

It is more difficult to calculate the false positive probability for the outer planet candidates. Because the orbital period is unconstrained, a \vespa-like false positive analysis loses an important piece of information (namely, the duration of the transit compared to the orbital period). Even though we can estimate the orbital period of the outer three planets (assuming they indeed transit \thisstar), we have no constraint on orbital periods for the scenario where the single transit signals are astrophysical false positives. We do, however, know that the three single-transits are detected in a multi-transiting planet candidate system and can use this fact to estimate the false positive probabilities without any knowledge of the transit shapes and orbital periods. \citet{lissauer} give expressions for estimating the likelihood of false positive signals in multiple planet systems. Using these expressions with numbers from the recent Data Release 24 \Kepler\ planet candidate catalog \citep{coughlin}, we find that the probability of a given target having two planets and three false positives is roughly $10^{-12}$, the probability of the target having three planets and two false positives is roughly $10^{-9}$, and the probability of the target having four planets and one false positive is roughly $5\times10^{-7}$. From the observed number of systems with five or more transiting planets discovered by \Kepler, the probability of a star hosting such a system is roughly $18/198646$ or $10^{-4}$. When we compare these probabilities, we find that, {\em a priori}, it is $10^8$ times more likely \thisstar\ hosts five transiting planets than two planets and three false positives, $10^5$ times more likely \thisstar\ hosts five transiting planets than three planets and two false positives, and about $200$ times more likely that \thisstar\ hosts five transiting planets than four planets and one false positive. When this information is combined with the fact that the transits are u-shaped (consistent with small planets) rather than v-shaped (consistent with a background false positive), and our adaptive optics imaging which rules out many possible background contaminants, we have high confidence that all five candidates in the \thisstar\ system are genuine planets.

\subsection{Dynamics}
The richness of the \thisstar\ planetary system gives rise to questions about its dynamics and architecture. In this section, we aim to address and place constraints on the dynamical stability of the system and the resonance state of the inner two planets. The dynamical stability arguments we make in this section are useful for constraining the orbital periods of the outer two planets (which we do in Section \ref{periods}).

\subsubsection{Inner Planets}
\label{sec:inner}
We first considered the two inner planets, which both show multiple transit events in the K2 light curve and therefore have precisely measured orbital periods. The ratio of the orbital periods of the two inner planets is just 3.57\% larger than 2:1, so we tested whether the two planets orbit in a 2:1 mean motion resonance (MMR).  

We assessed the resonant state of the inner two planets by conducting 10,000 numerical simulations of the orbits of the inner two planets over 100,000 years using the \texttt{Mercury6} N-body integrator \citep{mercury6}. For each trial, we drew the orbital elements of each planet from the posterior probability distribution from the MCMC transit fit (Section \ref{sec:lc_analysis}). We assigned masses to the planets using the the methodology of \citet{inc-oscillations} -- to summarize, given the planet radii, we draw masses from several published mass-radius relations: the \citet{weissmarcy} relation for planets with $R_{p} <1.5 R_{\oplus}$, the \citet{angie} relation for planets with $4 R_{\oplus} > R_{p} > 1.5 R_{\oplus}$, and for planets larger than 4\rearth\, we solve for mass by drawing the mean planetary density from a normal distribution centered at $\rho = 1.3 \pm 0.5$ g / cm$^{3}$, taking the hot Jupiter radius anomaly into account using the relation from \citet{laughlin-radanom}.  

We tested each of the 10,000 realizations of the system for resonant behavior. The condition for resonance is more stringent than that of a period commensurability: for a pair of planets to be resonant, they must have oscillating (rather than circulating) resonance angles, which means that the longitude of conjunction (the location where the the planets pass closest together) has an approximately stable location. Resonances are sometimes referred to as a ``bound states'' because planets can be trapped in the energetically favorable configuration where the resonance angles oscillate back and forth in a potential well, like a pendulum with an energy low enough to swing back and forth rather than swing 360 degrees over the top \citep{nodding}. At the same time, a pair of planets can have a period ratio slightly out of an integer ratio and still be in resonance. We examined the resonance argument of the inner two planets, $\varphi$, which is defined as:

\be
\varphi = (p + q) \lambda_{inner} - p \lambda_{outer} - q \varpi_{outer},
\label{eq:resarg}
\ee
where $p/(p+q)$ is the order of the resonance (which is 2:1, so $p=1$ and $q=1$ for these planets), $\varpi$ is longitude of pericenter, and $\lambda$ is angular location in orbit.

Out of the 10,000 system realizations that we tested, none of them were in resonance (all had circulating rather than oscillating resonant arguments). Therefore, we conclude that the inner two planets orbiting \thisstar\ do not orbit in a mean motion resonance. This conclusion is not surprising -- the sample of multi-planet systems from \Kepler\  shows that planets more often orbit near, but not in, mean motion resonances \citep{veras-res}. The fact that these planets orbit slightly outside of mean motion resonance is also reminiscent of trends seen in \Kepler\ multi-planet systems. \citet{fab2014} found that period ratios slightly larger than 2:1 (as is the case for these two planets) are overrepresented in the population of observed systems, and slightly smaller ratios are underrepresented. Thus, there is no evidence to suggest that these planets are in resonance, but they are a part of the overabundance of planets that pile up slightly outside the 2:1 MMR. 

We note that in this analysis, we have examined only the behavior of the two inner planets. The three outer planets in the system contribute additional terms in Equation \ref{eq:resarg}, which we have ignored because of their poorly constrained orbits, but which could presumably alter the resonant behavior of the inner two planets. However, we believe it is unlikely that the outer planets would significantly affect the inner planets' resonant state. The periods of the outer three planets are likely significantly longer (by an order of magnitude or so, see Section \ref{periods}) than the periods of the inner two planets, so the outer planets will act like distant static perturbers.

\subsubsection{Outer Planets}\label{outerplanets}

We performed a separate dynamical analysis to study possible orbits and configurations of the outer three planets in the \thisstar\ system. The outer three planets only transited \thisstar\ once during the 75 days of K2 observations, so their orbital periods are not uniquely determined from the light curve. We do, however, measure the transit duration, radius ratio, and impact parameter of the three single-transit events, and our follow-up spectroscopy and analysis measures the mean stellar density, which allows us to estimate the semi-major axes and orbital periods of the three outer planets (see Table \ref{bigtable} for the best-fit values for each parameter).

We assessed the dynamical stability of the system by performing 4000 N-body simulations using the \texttt{Mercury6} hybrid integrator. We initialized the N-body simulations with orbital elements drawn from either the posterior probability distributions of transit parameters or from reasonable priors. 

We estimated the outer singly-transiting planets' orbital periods (and therefore semimajor axes) from the transit and stellar parameters using an analytical expression \citep[e.g.,][]{seager2003} with a correction for nonzero eccentricity \citep[as in][]{duration-ford}:
\begin{multline} \label{eq:duration}
t_{d,i} = \frac{P_{i}}{\pi} \arcsin{}\left[\left(\frac{G (M_{*} + m_{p,i}) P_{i}^{2}}{4 \pi^{2}}\right)^{-1/3} \times \right. \\
\left. \sqrt{(R_{P,i} + R_{*})^2 - (b_{i}^2 \times R_{*}^2)} \right] \frac{\sqrt{1-e_{i}^2}}{1+e_{i} \cos{\varpi_{i}}}
\end{multline}
where $t_{d,i}$ is the transit duration of the $i$th planet (from first to fourth contact), $P_{i}$ is its period (for which we would like to solve),  $m_{p,i}$ is the mass of the $i$th planet, $e_i$ is the orbital eccentricity, $\varpi_i$ is the argument of periastron, $b_i$ is the impact parameter, $M_{*}$ is the stellar mass, $R_{*}$ is the stellar radius, and $G$ is the gravitational constant. 

We solved Equation (\ref{eq:duration}) numerically 4000 times for each outer planets' orbital period (which we then converted to semimajor axis). For each of the 4000 realizations, we drew the quantities $t_{d,i}$, $P_i$, $R_{P,i}$, and $b_i$ from the light curve posterior probability distributions from the MCMC transit fits. We generated the planet masses $m_{p,i}$ from the measured planet radii using the same piecewise mass-radius relation as was used in \ref{sec:inner}. Values for $M_{*}$ and $R_{*}$ were drawn from the posterior probability distributions generated in Section \ref{sec:stellar_prop}, and values for $e$ were drawn from a beta distribution with shape parameters $\alpha = 0.867$ and $\beta = 3.03$ \citep[derived from the population of observed planets given in][]{kipping-prior1, kipping-prior2, kipping16}. We used an asymmetric prior for the argument of periastron $\varpi$ to account for the fact that the planet is observed to be transiting \citep[the value of which is dependent on the drawn eccentricity; see Equation 19 in][]{kipping16}.

After determining initial parameters, we integrated each of the 4000 systems forward in time for 1 Myr, long enough to examine interactions over many secular periods, while requiring energy be conserved to one part in $10^{8}$. Of the total 4000 realizations, only a subset (roughly 10\%) were dynamically stable over 1 Myr timescales, meaning that dynamical arguments can help constrain the system architecture, including the orbits of the three outer planets. We found that the most important variables for determining the stability of the system are the orbital eccentricities of the individual planets. For a given transit duration, the orbital period and eccentricity are degenerate.  
As a result, the eccentricities' constraints translate into limits on the orbital periods of the outer planets. 

\begin{figure*} 
   \centering
   \includegraphics[width=6.5in]{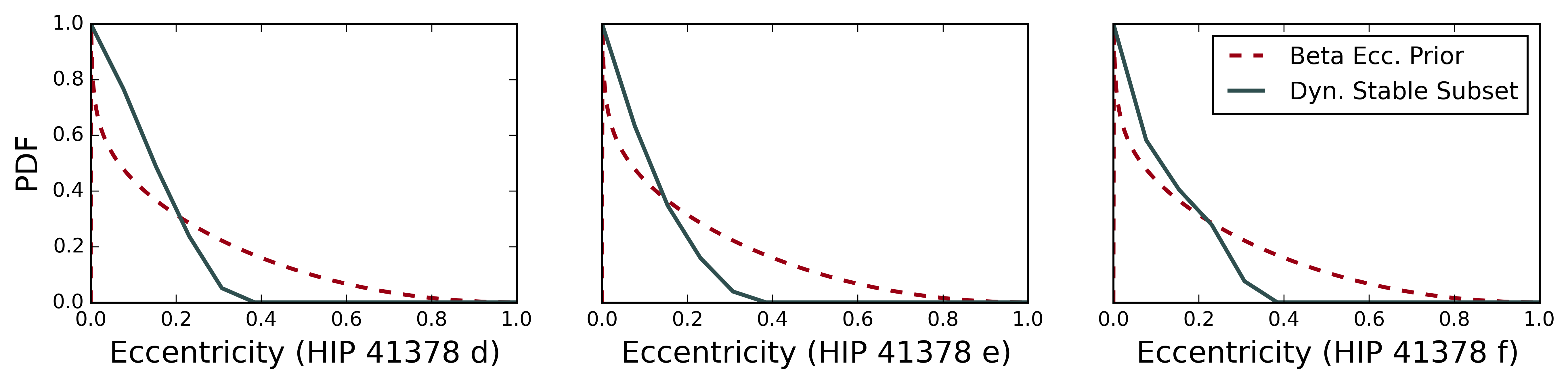} 
   \caption{Comparison of input planet eccentricity to dynamical simulations (red dashed lines) and the eccentricity of planets in dynamically stable systems (black solid lines). The input to the dynamical simulations is the distribution of eccentricities in all exoplanets detected with radial velocities \citep{kipping-prior1}. The difference in shape between the two curves demonstrates which eccentricities are preferred in dynamically stable systems. Evidently planets with eccentricities larger than $e\sim 0.37$ or so will cause the system to go dynamically unstable. The maximum of each curve is normalized to one to show the difference in shape between the two distributions.}
   \label{fig:stable_ecc_limits}
\end{figure*}

Figure \ref{fig:stable_ecc_limits} shows the difference in initial eccentricity distribution (namely, the beta distribution prior) and the eccentricity distribution of the planets in systems that remained dynamically stable. These distributions are visibly different, and the eccentricities of stable systems are preferentially lower. Among our 4000 realizations, all systems containing planets with eccentricities above $e\sim 0.37$ became dynamically unstable, suggesting that the true eccentricities are less than this value. 

\subsection{Orbital Periods of the Outer Planets} \label{periods}

In this section, we estimate the orbital period of the three outer planets transiting \thisstar\ under various assumptions and taking different information into account. We calculate orbital periods with a similar analysis to that described in the previous section, in particular by solving Equation \ref{eq:duration} numerically after drawing parameters from the MCMC transit fit posterior probability distributions or from priors. 

\begin{figure*} 
   \centering
   \includegraphics[width=6.5in]{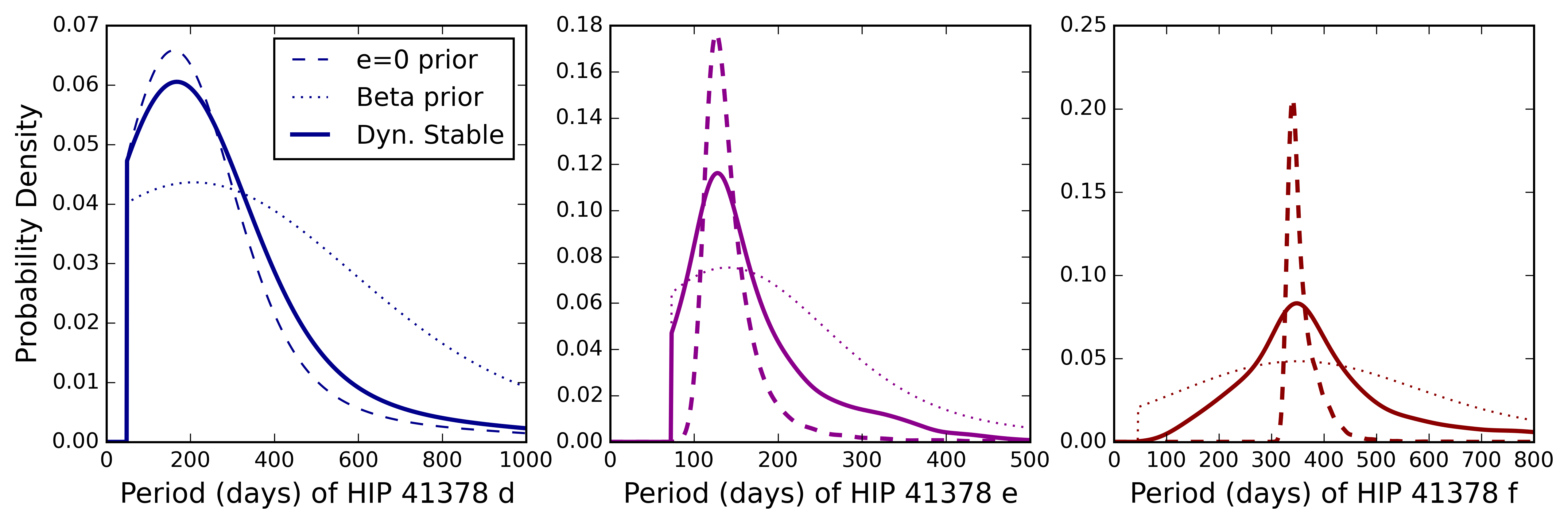} 
   \caption{Probability distributions for the orbital period of each of the outer planets in the system (detected by only a single transit in K2 data).  The dashed lines used a prior of null eccentricity for all three planets. The dotted lines used the Kipping beta distribution as the prior for eccentricity, with the prior for $\varpi$ being that from \citet{kipping16}, which accounts for both geometrical and observational biases. The solid lines use the Kipping eccentricity and $\varpi$ priors, but impose two additional priors of dynamical stability and transit probability. The area under each curve is normalized to one for ease of comparison.  }
   \label{fig:e0_dist}
\end{figure*}

We first calculated orbital periods under the assumption of strictly circular orbits. We also required that the orbital periods be longer than the baseline of K2 observations before and after each event -- otherwise, we would have seen multiple transits. We find that when we assume a circular orbit, we obtain relatively tight limits on the periods of the outer three planets, in particular the two deepest transits with precisely measured durations and impact parameters. For long period planets like these, however, the assumption of a circular orbit is in general not justified, so we believe these orbital period estimates are artificially tight. The distributions of orbital periods for the outer three planets assuming circular orbits are shown in Figure \ref{fig:e0_dist}. 

We also calculated orbital periods with the assumption of circular orbits relaxed to allow orbital eccentricities and arguments of periastron drawn from the same beta distribution and asymmetric prior described in the previous section (and which we used as an input to the dynamical simulations). As noted previously, this distribution matches the observed distribution of orbital eccentricity for exoplanets detected by radial velocities. When we do not assume circular orbits, the limits on the orbital periods are much looser. Although the median orbital periods we derived under the assumption of circular orbits and eccentric orbits are relatively similar, the width of the distribution changes drastically. In the case of \thisfifthplanet, the uncertainty on the orbital period increases by an order of magnitude when taking into account nonzero eccentricity. The orbital period distributions given this prior on orbital eccentricity are also shown in Figure \ref{fig:e0_dist}, where they can be compared to the case of circular orbits. 

The fact that the eccentricity of exoplanets tends to follow a beta distribution is not the only information we have about the system architecture or orbital eccentricities of the outer three planets. We can place additional constraints on the orbital eccentricities (and therefore orbital periods) by requiring that the system be dynamically stable. In Section \ref{outerplanets}, we found that the \thisstar\ system is dynamically unstable on 1 Myr timescales when any of the planets have eccentricities greater than 0.37, so we remove all orbital periods with eccentricities greater than 0.37. We also remove all systems that are not Hill stable \citep[using the criterion from ][]{fabrycky}. Enforcing these dynamical stability criteria narrow the distributions of plausible orbital periods by about 30\%. The orbital period distributions with dynamical stability enforced are also shown in Figure \ref{fig:e0_dist}, along with distributions without dynamical stability enforced for comparison. 

\begin{figure*} 
   \centering
   \includegraphics[width=6.5in]{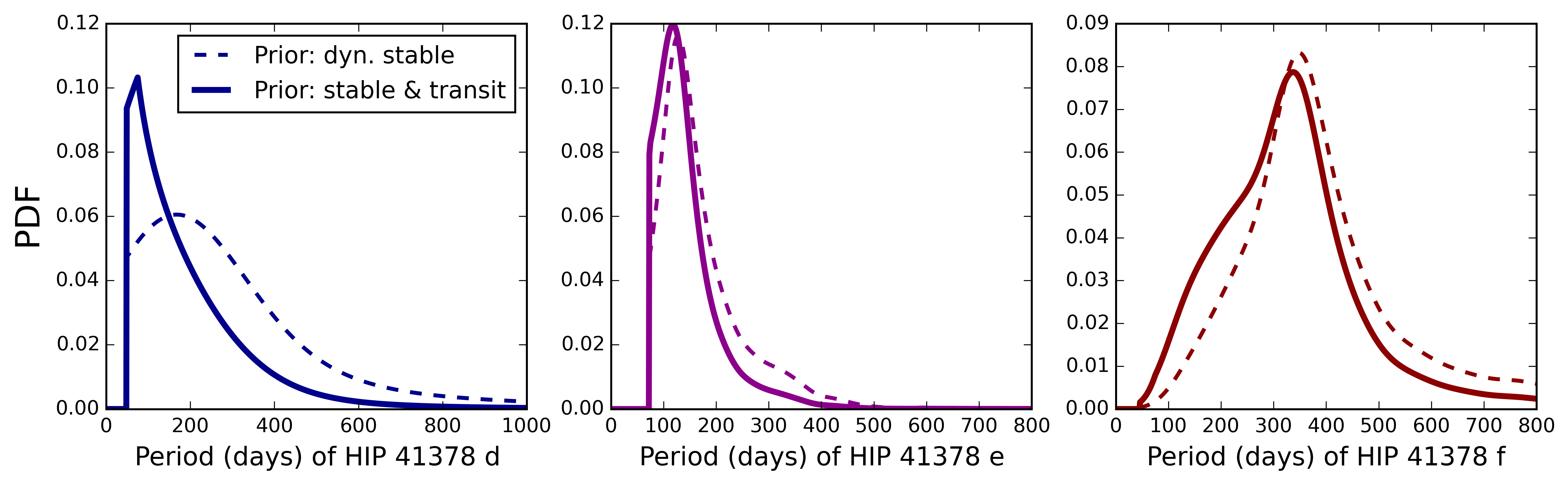} 
   \caption{Probability distributions for the orbital period of each of the single-transit planets in the system, incorporating dynamical stability alone (dashed lines) and incorporating dynamical stability and the probability of detecting a single transit with K2 (solid lines). The distribution only taking into account dynamical stability is the same as the solid lines shown in Figure \ref{fig:e0_dist}. Incorporating the prior information that these three planets transited during K2 observations sharpens our predictions of the orbital periods of the three outer planets.}
   \label{fig:dyn_final}
\end{figure*}

Finally, we took into account the fact that we observed these three planets to be transiting during the 75 days of K2 observations. Planets with shorter orbital periods are more likely to transit during a limited baseline than planets with longer orbital periods. We take this information into account by imposing a prior of the form:
\begin{equation}
\mathcal{P}(P_{i}, t_{d,i}, B)=
\begin{cases}
1 \ \ \ {\rm if}\ P_{i} - t_{d,i} < B \\
(B + t_{d,i}) / P_{i} \ \ \ {\rm else},\ \\
\end{cases}
\label{transitlhood}
\end{equation}
\noindent where $\mathcal{P}$ is the probability of observing a transit of planet $i$, $B$ is the time baseline of the observations, $t_{d,i}$ is the $i$th planet's transit duration, and $P_{i}$ is the orbital period of the planet $i$. Here, we define the planet being `observed to transit' as any part of its ingress or egress occurring during K2 observations. We imposed this prior on the orbital period distribution taking into account nonzero orbital eccentricity and dynamical stability, and we show the result in Figure \ref{fig:dyn_final}. The effect of this prior is to narrow the period distributions by another $\sim$30\% and to shift the period distributions to slightly lower values. The effect is most pronounced on the period distribution of \thisthirdplanet\, which had a weakly constrained orbital period because of its shallow transit. 

We summarize our orbital period estimates under these various assumptions in Table \ref{tab:per_lim}. We report the median values and $1\sigma$ widths of each distribution. In this paper, we choose to adopt the period distributions which were calculated taking into account nonzero eccentricity, dynamical stability, and the fact that the planets transiting during the K2 observations as our best estimate for the outer planets' orbital periods. These distributions incorporate the most information we have about the system to give the best possible period estimates.

\begin{deluxetable*}{lccr}
\tablewidth{0pt}
\tablehead{
\colhead{Eccentricity Prior} & \colhead{\thisthirdplanet\ period} &\colhead{\thisfourthplanet\ period} &\colhead{\thisfifthplanet\ period} }
\startdata
  $e=0$      & $157\substack{+195 \\ -41}$ & $132\substack{+37 \\ -14}$ & $348\substack{+37 \\ -13}$ \\
  $e$ beta distribution (as used in Section \ref{outerplanets})       & $188\substack{+397 \\ -87}$ & $143\substack{+129 \\ -52}$ & $367\substack{+311 \\ -130}$ \\
  $e$ beta distribution, dynamically stable only       & $174\substack{+260 \\ -68}$ & $140\substack{+92 \\ -43}$ & $361\substack{+182 \\ -103}$ \\
  \textbf{Adopted:} $e$ beta distribution, dyn. stable only + baseline prior      & $156\substack{+163 \\ -78}$ & $131\substack{+61 \\ -36}$ & $324\substack{+121 \\ -127}$ \\
\enddata
\tablecomments{Estimated periods for the three outer planets using four choices of priors. The $e=0$ prior produces the smallest errors on period, but it is likely these are underestimated. We adopt the results from the fourth line, which uses a the beta distribution for eccentricity and incorporates priors accounting for dynamical stability and transit likelihood (Equation \ref{transitlhood}) as our best estimates of the orbital periods in this system.} 
\label{tab:per_lim}
\end{deluxetable*}

\section{Discussion}
\label{discussion}

\begin{figure*} 
   \centering
   \includegraphics[width=6.5in]{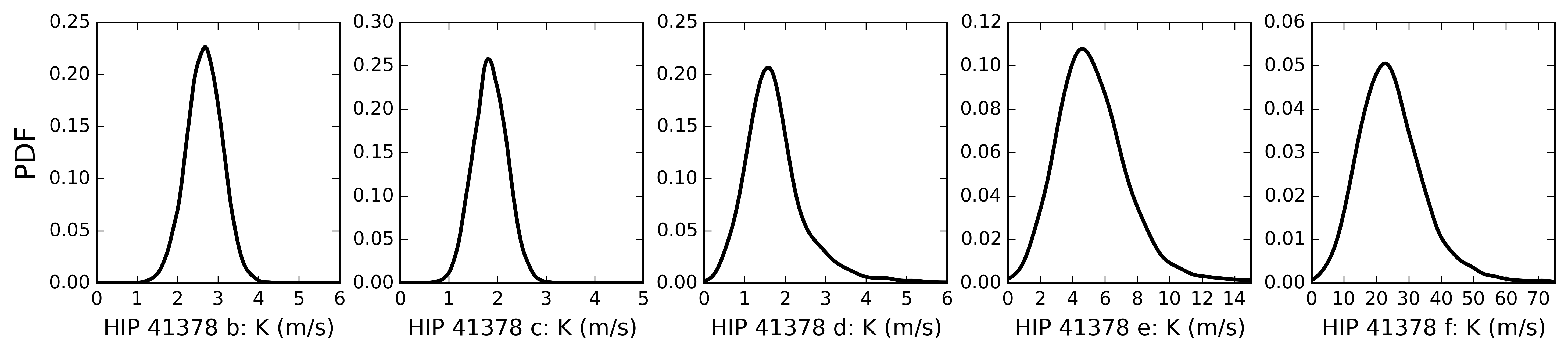} 
   \caption{Probability density function for the expected stellar reflex velocities caused by the motion of each planet in this system. Planets b and c have well measured orbital periods and ephemerides which will make it easier to measure their masses despite the low amplitudes of their RV signals.}
   \label{fig:kamp}
\end{figure*}

\thisstar\ is a compelling candidate for follow-up observations due to its brightness, the accessible size of the planets, and the system's rich architecture. \thisstar\ is the second brightest multi-transiting system, behind \Kepler-444 \citep{campante}, a system of five sub-Earth sized planets with expected RV semi-amplitudes below the noise-floor of current instrumentation. Unlike the \Kepler-444 system, the planets orbiting \thisstar\ should each have measurable RV semi-amplitudes. We have estimated the likely range of RV semi-amplitudes for each planet assuming planetary masses drawn from the \citet{angie} distribution and periods and eccentricities drawn from our analysis in Sections \ref{outerplanets} and \ref{periods}. The RV semi-amplitude distributions, shown in Figure \ref{fig:kamp}, are all centered above 1 \ms, and could therefore be detectable with spectrographs like HARPS-N \citep{harpsn} and HIRES \citep{hires} in the north, and HARPS \citep{harps} and PFS \citep{pfs} in the south. It will be most challenging to detect \thisthirdplanet, which has an unknown period (unlike the inner two planets) and most likely induces an RV semiamplitude of only 2 \ms, but such signals have been detected previously in intensive observing campaigns \citep[e.g.][]{lovis}.  

Radial velocity measurements will be particularly valuable for two reasons. First, precise mass measurements of the inner two planets can probe the mass radius diagram in the regime of low incident flux. Most transiting planets with precise masses orbit very close to their host stars, where any gaseous envelopes originally present might have been stripped by the intense stellar radiation. Planet masses measured from transit timing variations have shown that some planets on longer periods are likely less dense than most short period planets. Measuring precise masses of planets in longer period orbits (like the inner two planets in this system) can help show whether or not a planet's radiation environment affects its density. 

RV measurements will also be important to determine the orbital period of the outer planets, in particular the long-period gas giant \thisfifthplanet. This planet's long orbital period and the brightness of the host star make \thisfifthplanet\ a promising target for future transit follow-up studies -- provided a precise orbital period and transit ephemeris can be recovered. \thisstar\ is five magnitudes brighter than the recently discovered transiting Jupiter analog host \Kepler-168 \citep{kippingjupiteranalog}, making \thisfifthplanet\ one of the best long-period transiting planets for transit transmission spectroscopy. Studying \thisfifthplanet's atmosphere will open up a new regime for atmospheric studies, which typically focus on short period, highly irradiated planets. 

\thisfifthplanet\ could also be a compelling target to measure the planet's oblateness. The planets in our own solar system are not spherical and are distorted into oblate spheroids by the planets' rotation. A planet's projected oblateness can be measured from its transit light curve \citep{huiseager}. Indeed, strong constraints have been placed on the degree of oblateness for hot Jupiters \citep{carteroblate,zhu}, even though these planets are not expected to show measurable oblateness because their rotation periods are likely to be synchronized with their orbits \citep{seagerhui}. The long period of \thisfifthplanet\ implies that its rotation will not have tidally synchronized with its orbit, so its oblateness is likely to be large enough to detect. 

Follow-up observations of \thisfifthplanet\ hinge on our ability to recover a precise orbital period and transit ephemeris. Because of the planet's (apparent) long orbital period, it may be difficult to measure the spectroscopic orbit precisely enough to recover transits using a non-dedicated instrument like \Spitzer. The CHEOPS spacecraft \citep{broeg} may be the ideal instrument to recover transit ephemerides for the outer planets (and therefore precise orbital periods) given the mission focus on transiting planets. Long-term monitoring with CHEOPS may also reveal additional transiting planet candidates with long orbital periods which did not happen to transit during the K2 observations. 

The challenges we face attempting to measure precise orbital periods for planets with just a single transit in the \thisstar\ light curve are not unique to this system. Numerous single-transits have been observed in both \Kepler\ data \citep{wang} and K2 data \citep{hip116454,osborn}. Previous studies \citep{yee, wang, osborn} have shown that it is possible to make sharp predictions of the orbital period of a planet with a single transit, assuming strictly circular orbits. This assumption is in general not justified for long-period exoplanets, where radial velocity surveys have shown that high eccentricities (greater than those observed in our solar system) are common \citep{kipping-prior1}, and weakly constrained eccentricity substantially increases the range of possible orbital periods \citep{yee}. Here, we are able to take orbital eccentricity into account and still obtain relatively strong constraints on orbital periods by incorporating priors on eccentricity (from RV planet detections), dynamically stability, and detecting transits during the timespan of K2 observations. These techniques should be developed further. In future work, they will provide valuable tools for estimating periods and other orbital elements for singly-transiting planets in multi-planet systems from \Kepler, K2, or TESS \citep{ricker}, which will observe most of the sky for only 28 days and will likely discover over 100 planets with a single transit event \citep{sullivan}. 

\section{Summary}
\label{summary} 

Using data from the K2 mission, we have discovered, validated, and characterized the \thisstar\ planetary system. Our main results can be summarized as follows: 

\begin{enumerate}
\item \thisstar\ hosts a system of at least five transiting planets, three of which were discovered by observing (only) a single transit. The two inner planets, in 15.6 and 31.7 day orbits, have radii $R_P$ = 2.9 and 2.6 \rearth\ respectively. The three single-transit planets have radii $R_P$ = 4, 5.5, and 10 \rearth, and orbital periods likely longer than 100 days. These planets orbit a particularly bright F-type star, \thisstar. The host \thisstar\ is a slightly evolved F-star with a V-band magnitude of 8.9, an H-band magnitude of 7.8, and a K-band magnitude of 7.7. As a result, its planetary system is a good candidate for follow-up observations.

\item The outer three planets only transited \thisstar\ once during the 75 days of K2 observations. Although orbital periods are not well-defined for single transit events, we have constrained the orbital periods of the newly discovered planets. Using our knowledge of the host's stellar properties and the planets' transit parameters, and a reasonable prior on the orbital eccentricity, we estimate a range of possible orbital periods for the outer three planets. We are able to sharpen these estimates by a factor of two by incorporating information about the system's dynamical stability and the probability of a transit being observed during the K2 observations. We find that the most likely periods for the three new planets are \perpld, \perple, and \perplf\ days for planets d, e, and f, respectively. 

\item Follow-up radial velocity observations could measure masses for all of the planets and could determine orbital periods for the three outer planets. We calculate that the reflex velocities on \thisstar\ from \thisfirstplanet, \thissecondplanet, and \thisthirdplanet\ are likely to fall in the range  2 -- 4 \ms, and thus are detectable with current instrumentation. The two inner planets have known periods, which will aid in isolating the RV signals of the outer planets. \thisfourthplanet\ and \thisfifthplanet\ have expected reflex velocities of approximately 5 and 25 m/s, respectively, and should be readily detectable with enough observational coverage.

\item \thisfifthplanet\ is a gas giant in a likely 1 year orbit. The host star's brightness and \thisfifthplanet's 0.5\% transit depth make it an attractive target for future transit follow-up observations if its precise orbital period and transit ephemeris can be recovered. \thisfifthplanet\ is one of the first gas giants with a cool equilibrium temperature transiting a star bright enough for transit transmission spectroscopy. It could also be possible to measure the planet's oblateness, since its orbital period is long enough that its rotation will not have synchronized with its orbit. 
\end{enumerate}

Discoveries such as the \thisstar\ system show the value of wide-field space-based transit surveys. The \Kepler\ spacecraft had to search almost 800 square degrees of sky (or seven fields of view) before finding such a bright multi-planet system suitable for follow-up observations. \thisstar\ is a preview of the type of discoveries the all-sky TESS survey will make routine.

\acknowledgments
We thank the anonymous referee for helpful comments on the manuscript. A.V. and J.C.B are supported by the NSF Graduate Research Fellowship, Grants No. DGE 1144152 and DGE 1256260, respectively. D.W.L. acknowledges partial support from the Kepler mission under NASA Cooperative Agreement NNX13AB58A with the Smithsonian Astrophysical Observatory. C.B. acknowledges support from the Alfred P. Sloan Foundation. 

This research has made use of NASA's Astrophysics Data System and the NASA Exoplanet Archive, which is operated by the California Institute of Technology, under contract with the National Aeronautics and Space Administration under the Exoplanet Exploration Program. This work used the Extreme Science and Engineering Discovery Environment (XSEDE), which is supported by National Science Foundation grant number ACI-1053575. This research was done using resources provided by the Open Science Grid, which is supported by the National Science Foundation and the U.S. Department of Energy's Office of Science. The National Geographic Society--Palomar Observatory Sky Atlas (POSS-I) was made by the California Institute of Technology with grants from the National Geographic Society. The Oschin Schmidt Telescope is operated by the California Institute of Technology and Palomar Observatory.

This paper includes data collected by the \Kepler\ mission. Funding for the \Kepler\ mission is provided by the NASA Science Mission directorate. Some of the data presented in this paper were obtained from the Mikulski Archive for Space Telescopes (MAST). STScI is operated by the Association of Universities for Research in Astronomy, Inc., under NASA contract NAS5--26555. Support for MAST for non--HST data is provided by the NASA Office of Space Science via grant NNX13AC07G and by other grants and contracts.

Robo-AO KP is a partnership between the California Institute of Technology, University of Hawai'i Manoa, University of North Carolina, Chapel Hill, the Inter-University Centre for Astronomy and Astrophysics, and the National Central University, Taiwan.  Robo-AO KP was supported by a grant from Sudha Murty, Narayan Murthy, and Rohan Murty.  The Robo-AO instrument was developed with support from the National Science Foundation under grants AST-0906060, AST-0960343, and AST-1207891, the Mt. Cuba Astronomical Foundation, and by a gift from Samuel Oschin.  Based in part on observations at Kitt Peak National Observatory, National Optical Astronomy Observatory (NOAO Prop. ID: 15B-3001), which is operated by the Association of Universities for Research in Astronomy (AURA) under cooperative agreement with the National Science Foundation. 

Facilities: \facility{Kepler/K2, FLWO:1.5m (TRES), KPNO:2.1m (Robo-AO)}


\begin{deluxetable*}{lcccc}
\tablecaption{System Parameters for \thisstar \label{bigtable}}
\tablewidth{0pt}
\tablehead{
  \colhead{Parameter} & 
  \colhead{Value}     &
  \colhead{} &
  \colhead{68.3\% Confidence}     &
  \colhead{Comment}   \\
  \colhead{} & 
  \colhead{}     &
  \colhead{} &
  \colhead{Interval Width}     &
  \colhead{}  
}
\startdata
\emph{Stellar Parameters} & & & \\
Right Ascension & 8:26:27.85 & & &  \\
Declination & +10:04:49.35 & & &  \\
Distance to Star~[pc]& 116 &$\pm$&18& A\\
V magnitude & 8.93 &  & & A\\ 
$M_\star$~[$M_\odot$] & \mstar & $\pm$&$ \mstare$ & C \\
$R_\star$~[$R_\odot$] & \rstar & $\pm$&$ \rstare$ & C \\
Limb darkening $q_1$~ & \ldone  & $\pm$&$ \uldone$ & D \\
Limb darkening $q_2$~ & \ldtwo  & $\pm$&$ \uldtwo$ & D \\

$\log g_\star$~[cgs] & \logg & $\pm$& \logge & C \\
Metallicity \metallicity & \mh & $\pm$&\mhe & B \\
$T_{\rm eff}$ [K] & \teff & $\pm$&$ \teffe$ & B\\
 & & \\
 
\emph{\thisfirstplanet} & & & \\
Orbital Period, $P$~[days] & \perplb & $\pm$&$ \uperplb $ & D \\
Radius Ratio, $(R_P/R_\star)$ & \rprstb & $\pm$&$ \urprstb$ & D \\
Scaled semimajor axis, $a/R_\star$  & \arstb & $\pm$&$ \uarstb$ & D \\
Orbital inclination, $i$~[deg] & \inclb & $\pm$&$ \uinclb$ & D \\
Transit impact parameter, $b$ & \impb & $\pm$&$ \uimpb$ & D \\
Time of Transit $t_{t}$~[BJD] & \ttransitb & $\pm$& \uttransitb & D\\ 
$R_P$~[\rearth] & \rplb &   $\pm$&$ \urplb$  & C,D \\
 & & \\
 
\emph{\thissecondplanet} & & & \\
Orbital Period, $P$~[days] & \perplc & $\pm$&$ \uperplc $ & D \\
Radius Ratio, $(R_P/R_\star)$ & \rprstc & $\pm$&$ \urprstc$ & D \\
Scaled semimajor axis, $a/R_\star$  & \arstc & $\pm$&$ \uarstc$ & D \\
Orbital Inclination, $i$~[deg] & \inclc & $\pm$&$ \uinclc$ & D \\
Transit Impact parameter, $b$ & \impc & $\pm$&$ \uimpc$ & D \\
Time of Transit $t_{t}$~[BJD] & \ttransitc & $\pm$& \uttransitc & D\\ 
$R_P$~[\rearth] & \rplc &   $\pm$&$ \urplc$  & C,D \\
 & & \\

\emph{\thisthirdplanet} & & & \\
Orbital Period, $P$~[days] & \perpld & & & E \\
Radius Ratio, $(R_P/R_\star)$ & \rprstd & $\pm$&$ \urprstd$ & D \\
Transit Impact Parameter, $b$ & \impd & $\pm$&$ \uimpd$ & D \\
Time of Transit $t_{t}$~[BJD] & \ttransitd & $\pm$& \uttransitd & D\\ 
Transit Duration $D$~[hours] & \tdurd & $\pm$& \utdurd & D\\ 
$R_P$~[\rearth] & \rpld &   $\pm$&$ \urpld$  & C,D \\
 & & \\
 
\emph{\thisfourthplanet} & & & \\
Orbital Period, $P$~[days] & \perple & & & E \\
Radius Ratio, $(R_P/R_\star)$ & \rprste & $\pm$&$ \urprste$ & D \\
Transit Impact Parameter, $b$ & \impe & $\pm$&$ \uimpe$ & D \\
Time of Transit $t_{t}$~[BJD] & \ttransite & $\pm$& \uttransite & D\\ 
Transit Duration $D$~[hours] & \tdure & $\pm$& \utdure & D\\ 
$R_P$~[\rearth] & \rple &   $\pm$&$ \urple$  & C,D \\
 & & \\
 
\emph{\thisfifthplanet} & & & \\
Orbital Period, $P$~[days] & \perplf & & & E \\
Radius Ratio, $(R_P/R_\star)$ & \rprstf & $\pm$&$ \urprstf$ & D \\
Transit Impact Parameter, $b$ & \impf & $\pm$&$ \uimpf$ & D \\
Time of Transit $t_{t}$~[BJD] & \ttransitf & $\pm$& \uttransitf & D\\ 
Transit Duration $D$~[hours] & \tdurf & $\pm$& \utdurf & D\\ 
$R_P$~[\rearth] & \rplf &   $\pm$&$ \urplf$  & C,D \\
 & & \\

\enddata

\tablecomments{A: Parameters come from \hipparcos. B: Parameters come from spectroscopic analysis with SPC. C: Parameters come from interpolation of parallax, V-magnitude, metallicity, and effective temperature onto model isochrones D: Parameters come from analysis of the K2 light curve. E: Constraints on orbital periods for singly-transiting planets are drawn from the posterior probability distributions of the transit parameters and stellar density, with priors imposed on the eccentricity, dynamical stability, and detecting transits.}

\end{deluxetable*}
\clearpage

\end{document}